\documentclass[useAMS,usenatbib]{mn2e}

\usepackage[dvips]{graphicx}
\usepackage{wrapfig}
\usepackage{amsmath}

\title[Radiative Regulation of Population III Star Formation]
{Radiative Regulation of Population III Star Formation}
\author[K. Hasegawa, M. Umemura and H. Susa]
{K. Hasegawa $^{1}$\thanks{E-mail: hasegawa@ccs.tsukuba.ac.jp (KH); umemura@ccs.tsukuba.ac.jp (UM); susa@konan-u.ac.jp (HS)} 
, M. Umemura$^{1}$
and H. Susa$^{2}$
\footnotemark[1]\\
$^{1}$Center for Computational Sciences, University of Tsukuba, Ten-nodai, 1-1-1 Tsukuba, 
Ibaraki 305-8577, Japan\\
$^{2}$Department of Physics, Konan University, Japan}
\begin{document}

\date{Accepted 1988 December 15. Received 1988 December 14; in original form 1988 October 11}

\pagerange{\pageref{firstpage}--\pageref{lastpage}} \pubyear{2008}

\maketitle

\label{firstpage}

\begin{abstract}
We explore the impact of ultraviolet (UV) radiation from massive
Population III (Pop III) stars 
of 25, 40, 80, and 120 $M_\odot$ on the subsequent Pop III
star formation.
In this paper, particular attention is paid to
the dependence of radiative feedback on the mass of source Pop III star. 
UV radiation from the source star can work to impede
the secondary star formation through the photoheating and
 photodissociation processes.
Recently, Susa \& Umemura (2006) have shown that the ionizing radiation
alleviates the negative effect by  $\rm H_2$-dissociating radiation from
120$M_\odot$ PopIII star,
since an $\rm H_2$ shell formed ahead 
of an ionizing front can effectively shield $\rm H_2$-dissociating
 radiation.
On the other hand, it is expected that the negative feedback by
$\rm H_2$-dissociating radiation can be predominant if a source star is
less massive, since a ratio of the $\rm H_2$-dissociating photon number
to the ionizing photon number becomes higher.
In order to investigate the radiative feedback effects from such less
 massive stars,
we perform three-dimensional radiation hydrodynamic simulations,
incorporating the radiative transfer effect of ionizing and $\rm H_2$-dissociating
radiation.
As a result, we find that if a source star is less massive than 
$\approx 25M_{\odot}$, the ionizing radiation cannot suppress the negative 
feedback of $\rm H_2$-dissociating radiation.
Therefore, the fate of the neighboring clouds around such less massive
stars is determined solely by the flux of 
$\rm H_2$-dissociating radiation from source stars.
With making analytic estimates of $\rm H_2$ shell formation 
and its shielding effect, 
we derive the criteria for radiation hydrodynamic feedback
depending on the source star mass. 
\end{abstract}


\section{Introduction}
The reionization and metal enrichment of the universe 
are thought to begin with the formation of first metal-free (Pop III) stars 
\citep{Gnedin00,Ciardi01,Cen03,Sokasian04}.
Hence, the formation rate of Pop III stars is crucial 
for the subsequent structure formation in the universe.
The Pop III objects are expected to collapse at $20 \la z \la 30$, 
forming a minihalo with a mass of $\approx 10^6 M_\odot$ and 
an extent of $\approx 100$pc
\citep{Tegmark97,Fuller00,Yoshida03}.
In the course of bottom-up structure formation, 
such Pop III minihaloes merge to form first galaxies at $z \ga 10$,
having the virial temperature $T_{\rm vir} \ga 10^4$K and the mass $\ga 10^8M_\odot$.
Even in the evolution of first galaxies, Pop III stars can play a significant role, 
since an appreciable number of stars may form from metal-free 
component in interstellar gas \citep{Tornatore07,Johnson08}.

The formation of very first stars has been investigated intensively in the last
decade. Many studies have come to a similar conclusion that such stars
form in a top-heavy mass function with the peak of $\approx 100 M_\odot$
\citep[e.g.,][]{Abel00, Bromm02, NU01, Yoshida06}. 
Recently, \citet{Oshea07} have shown that the variations
of cosmological density fluctuations allow the mass of Pop III stars 
to be down to $\sim 20M_\odot$. 

On the other hand,
the secondary Pop III star formation has been investigated recently.
The formation of secondary stars is subject to various feedback effects 
by first stars. One of them
is the supernova (SN) feedback through mechanical and chemical effects.
The negative feedback by SNe is the evaporation of neighboring
clouds, since the SN shock heats up the gas in clouds. 
On the hand, SNe can bring positive feedback through
the compression by shock and the cooling by
ejected heavy elements. The secondary star
formation can be promoted by such positive feedback effects
\citep{Mori02,Bromm03,KY05,Greif07}.
Another important feedback effect is
brought by the ultraviolet (UV) radiation from first stars,
since they are very luminous at ultraviolet band. 
First stars photoionize and photoheat the surrounding media,
and also photodissociate H$_2$ molecules, 
which are the main coolant of primordial gas.
The radiative feedback from first stars is the primary feedback
until first stars end the life-time of $\sim 10^6$yr 
with SNe.

The photodissociation of H$_2$ molecules leads to
a negative radiative feedback effect, which has been studied
by many authors so far. \citet{ON99} investigated the effect of
H$_2$-dissociating radiation from a single Pop III star residing 
in a virialized halo. 
They found that if the halo is uniform, H$_2$ molecules in the halo are
totally dissociated, so that the gas cannot collapse to form stars.
\citet{GB01} considered more realistic clumpy halos. 
They found that if the gas density is sufficiently high,
photodissociation process proceeds slower than the collapse of the
cloud. Hence, the cloud can form stars. This result is confirmed by
the recent 3D radiation hydrodynamic simulation by \citet{Susa07} including
the effects of hydrodynamics as well as the radiation transfer of
H$_2$-dissociating radiation. 
The feedback effects by diffuse
H$_2$-dissociating radiation can be important after the local
feedback in minihaloes
\citep{Haiman97,Macha01,Yoshida03,Oshea08}.

These works basically focused on the photodissociation effects. We also
have to take into account the effects of ionizing photons. Ionizing
radiation heats up the gas through the photoionization processes. 
The temperature of photoheated gas is kept to be around $10^4$K, 
owing to the balance between the radiative cooling and photoheating. 
If the gravitational potential of star forming halos are not so deep
as to retain the photoheated gas, the heated gas evaporates from the
halos \citep[e.g.,][]{SU04a,SU04b,Yoshida07b,WA08,Whalen08}.
However, the case in which ionizing radiation is coupled with
H$_2$-dissociating radiation is complex. 
When an ionization front (I-front) propagates in a collapsing core,
the enhanced fraction of electrons catalyzes H$_2$ formation 
\citep{SK87,KS92,Susa98,OH02}. In particular,
the mild ionization ahead of the I-front generates an H$_2$ shell,
which potentially shields H$_2$ dissociating photons \citep{Ricotti01}.
This mechanism is likely to work positively to form Pop III stars. 
On the other hand, the I-front can be accompanied with
a shock for an optically-thick cloud \citep{SU06}. 
The shock affects significantly the collapse of cloud. 
This is a totally radiation
hydrodynamic (RHD) process. Such radiation hydrodynamic feedback
has been investigated by 1D spherical RHD simulations 
\citep{AS07}, 2D cylindrical RHD simulations \citep{Whalen08},
and 3D RHD simulations \citep{SU06}. 
The results by 2D and 3D simulations are in good agreement with each other.
It is found that ionizing radiation can bring positive feedback
through the formation of H$_2$ shell. 

\citet{SU06} investigated RHD feedback 
by a $120M_\odot$ source star, and \citet{SU08} 
derived the feedback criterion. However,
if a source star is less massive, the relative intensity of H$_2$-dissociating 
radiation to ionizing radiation increases. Then, the feedback tends to be
more negative. 
In fact, the mass of first stars might be some $10M_\odot$
owing to the variations of cosmological density fluctuations \citep{Oshea07}, 
the enhanced H$_2$ cooling in pre-ionized gas \citep{SK87,Susa98,OH02},
or the HD cooling in fossil HII regions
\citep[e.g.,][]{UI00, NU02,Nagakura05,JB06,GB06,Yoshida07a}. 
Also, the elemental abundance patterns of hyper-metal-poor stars 
well match the yields by supernova explosions with a 
progenitor mass of $\sim 25M_{\odot}$ \citep{Umeda03, Iwamoto05}. 
The RHD feedback effects by Pop III stars less massive 
than $100M_\odot$ have not been investigated so far,
and no criterion has not been derived. 

In this paper, we perform 3D RHD simulations in
order to investigate the radiative feedback effects from Pop III stars
with various masses. We derive the criteria for
the collapse of cloud cores irradiated by a neighboring Pop III star
with 25,40,80, or  $120M_\odot$.
In \S2, the simulation code and procedure are described. 
The simulation results are presented in \S3. 
Finally, we summarize the conclusions in \S4. 

\begin{table}
  \caption{Properties of Pop III source stars}
  \begin{center}
  \begin{tabular}{llll}
  \hline
  Mass  & $T_{\rm eff}$ [K] & $\dot N_{\rm ion}$ [${\rm s^{-1}}$] & $L_{\rm LW}$[{\rm erg/s}] \\
  \hline
  $120M_{\odot}$ & $9.57\times 10^4$ & $1.069\times 10^{50}$ & $5.34 \times 10^{23}$ \\
  $80M_{\odot}$ & $9.33\times 10^4$ & $5.938\times 10^{49}$ & $3.05 \times 10^{23}$ \\
  $40M_{\odot}$ & $7.94\times 10^4$ & $1.873\times 10^{49}$ & $1.17 \times 10^{23}$ \\
  $25M_{\odot}$ & $7.08\times 10^4$ & $5.446\times 10^{48}$ & $3.94 \times 10^{22}$ \\
   \hline
\end{tabular}
\end{center}
\end{table}

\section[]{SIMULATION CODE AND PROCEDURE}\label{code}
We perform RHD simulations with a 3D Radiation-SPH code developed by ourselves. 
In the code, we treat self-consistently the gravitational force, 
hydrodynamics, the radiative transfer of UV photons, 
non-equilibrium chemistry  
for $\rm e^-$, $\rm H^+$, $\rm H$, 
$\rm H^-$, $\rm H_2$, and $\rm H_2^+$. 
We use the chemical network solver in \cite{Kitayama01} as well as 
the radiative transfer solver described in \cite{Susa06}. 
For the shielding by $\rm H_2$ molecules against H$_2$-dissociating 
radiation at Lyman-Werner (LW) band (11.26-13.6 eV), 
we employ the self-shielding function introduced by \cite{DB96}.
The opacity against LW band flux ($F_{\rm LW}$)  is 
calculated by
\begin{equation}
F_{\rm LW} = F_{\rm LW,0} f_{\rm s}
\left( N_{\rm H_2,14 } \right) \label{LW}
\end{equation}
where $ F_{\rm LW,0}$ is the incident flux, 
$ N_{\rm H_2,14}$ is the H$_2$ column density in units of 
$10^{14}{\rm cm^{-2}}$, and
\begin{equation}
f_{\rm s}(x) = \left\{
\begin{array}{cc}
1,~~~~~~~~~~~~~~x \le 1 &\\
x^{-3/4},~~~~~~~~~x > 1 &
\end{array}
\right. \label{shield-function}
\end{equation}

\begin{figure}
	\centering
	{\includegraphics[width=7cm]{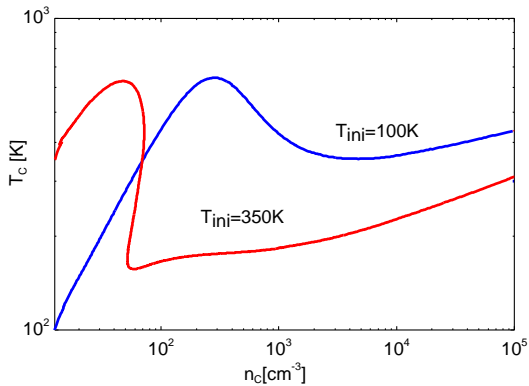}}
	\caption{Evolution of the core without radiative feedback. 
	The time variations of core temperature are shown 
    	for the initial temperature 350K (red curve) and 
	100K (blue curve), respectively. }
	\label{rhoT}
\end{figure}

In this paper, 
we simulate the evolution of a purely baryonic primordial cloud,
according with the model by \citet{SU06}. 
The cloud is initially uniform with the density of $n_{\rm H}=14 {\rm cm^{-3}}$, 
and has the mass of $M=8.3\times 10^4M_{\odot}$. 
The initial chemical compositions are assumed to be the cosmological 
compositions provided by \citet{GP98}.
Before the UV irradiation, the cloud contracts self-gravitationally 
to form a collapsing core. As for the core temperature $T_{\rm c}$, 
we employ two models.
One is a high temperature model,
and the other is a low temperature model.
By changing the initial temperature of 
the clouds $T_{\rm ini}$, we realize such core temperatures. 
As shown in Fig. \ref{rhoT}, if we set $T_{\rm ini}=100$K, 
the core temperature becomes $T_{\rm c}\sim 300-400$K at core density 
$n_{\rm c} \ga 10^2 {\rm cm^{-3}}$. 
On the other hand, 
if the initial temperature is set to be $T_{\rm ini}=350$K,
the cloud core cools below $T_{\rm c}\sim 200$K, 
since $\rm H_2$ molecules are rapidly formed 
owing to the high initial temperature (see Fig.\ref{rhoT}). 
Another difference between two models is the ratio
of gravitational energy $W$ to internal energy $U$, because
it is dependent on the initial temperature.
The $|W|/U$ ratio is $\approx 4$ for the low initial temperature 
(high $T_{\rm c}$) model,
while the $|W|/U$ ratio is $\approx 2$ for the high initial temperature 
(low $T_{\rm c}$) model.

We ignite a source star when the density of cloud core exceeds 
a certain value $n_{\rm on}$.
The source star is placed $D$ pc away from the center of cloud core. 
We change the mass of source star in the range of 
$25M_{\odot} \leq M_* \leq 120M_{\odot}$.
The properties of source stars as
the effective temperature of star $T_{\rm eff}$,
the number of ionizing photons emitted per second 
$\dot N_{\rm ion}$, and the luminosity at LW band 
are taken from \cite{Schaerer02}, which are summarized in Table 1.  
Note that we do not consider the lifetimes of source stars in this paper, 
since we focus on elucidating the RHD feedback before SN explosions. 

Numerical runs are characterized by 
the parameters $D$, $n_{\rm on}$, and $M_{*}$. 
The simulations are performed until $t_{\rm end}= 2t_{\rm ff}$, 
where $t_{\rm ff}$ is the free-fall time determined by $n_{\rm on}$. 
If the density of cloud core exceeds $5\times 10^5 {\rm cm^{-3}}$
before $t_{\rm end}$, we stop the calculation,
since the cloud is expected to keep collapsing. 
In order to clarify the effects of the ionizing radiation,
we also perform the simulations artificially 
disregarding ionizing radiation but still including LW radiation, 
and the results are compared with those of normal simulations. 
The number of SPH particles 
handled in our simulations is 262,144 for all runs. 

The present simulations are mainly carried out with 
a novel hybrid computer system in University of Tsukuba, 
called {\it FIRST} simulator, which has been designed to simulate 
multi-component self-gravitating radiation hydrodynamic systems
with high accuracy \citep{Umemura07}. 
The {\it FIRST} simulator is composed of 256 nodes with dual Xeon processors, 
and each node possesses a Blade-GRAPE board, on which GRAPE-6 chips,
that is, the accelerator of gravity calculations, are implemented. 
The peak performance of {\it FIRST} simulator is 36.1 Tflops.

\section{TYPICAL RESULTS}
\begin{figure}
	\centering
	{\includegraphics[width=7cm]{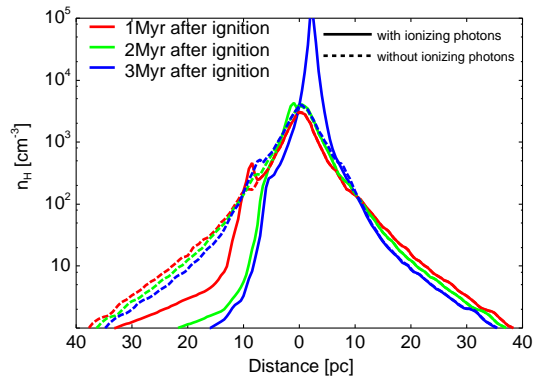}}
	\caption{Time variations of density profiles along the symmetry axis 
	for a simulation with 
    $T_{\rm c}\sim 400$K, $M_{*} = 80M_{\odot}$, 
    $n_{\rm on}=10^3{\rm cm^{-3}}$, and $D=40{\rm pc}$. 
	The red, green, and blue lines represent the profiles at 1Myr, 2Myr, and 
	3 Myr after the ignition, respectively. 
	The results without ionizing radiation are shown by dashed lines.
    It is shown that the core cannot collapse without ionizing radiation.}
	\label{evo80M}
\end{figure}
\begin{figure}
	\centering
	{\includegraphics[width=7cm]{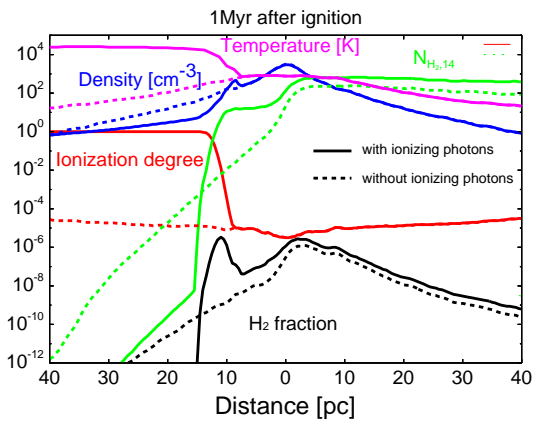}}
	\caption{Various physical quantities along the symmetry axis at 1Myr after the ignition 
	for the simulation shown in Fig. \ref{evo80M}. 
	The results with and without ionizing radiation are shown 
    	by solid and dashed lines, respectively.	
	The magenta, blue, red, green, and black lines 
	show the gas temperature(K), number density ($\rm cm^{-3}$), 
	electron fraction, $\rm H_2$ column density from the source star in units of 
	$10^{14}{\rm cm^{-2}}$, and the $\rm H_2$ fraction, respectively.}
	\label{vpq80M}
\end{figure}
In this section, we show the typical evolution of clouds.
For a high core temperature model ($T_{\rm c}\sim 300-400$K),
the time evolution of density profiles along the symmetry 
axis is shown in Fig. \ref{evo80M}, where the set-up parameters are
$M_{*} = 80M_{\odot}$, $n_{\rm on}=10^3{\rm cm^{-3}}$, and $D=40{\rm pc}$.
In this figure, the results with ionizing radiation are compared to
those without ionizing radiation.  
In the simulation with ionizing radiation, 
the density of cloud core keeps increasing, and the density
exceeds the limit ($5\times 10^5 {\rm cm^{-3}}$) due to the run-away collapse at 3.4Myr 
after the ignition of the source star. 
On the other hand, 
in the simulation without ionizing radiation, the gravitational
contraction of cloud core is stopped by the thermal pressure,
and eventually a hydrostatic core forms.

Various physical quantities along the symmetry axis at 1Myr 
are shown in Fig. \ref{vpq80M}. 
If ionizing radiation is included, a dense shell forms ahead of
the ionization front (I-front) . The $\rm H_2$ molecule fraction is raised up to
a level $y_{\rm H_2} \approx 10^{-5}$ in the shell, so that
$\rm H_2$ column density exceeds
$10^{14}{\rm cm^{-2}}$. 
Owing to the self-shielding of LW band radiation by the shell,
the $\rm H_2$ fraction in the cloud core is increased, compared to
the case without ionizing radiation.  
Eventually, the enhanced $\rm H_2$ cooling allows the
core to undergo the run-away collapse. 
On the other hand, unless ionizing radiation is included, 
LW band radiation from the source star reduces the $\rm H_2$ fraction,
so that the cloud core is settled in a hydrostatic configuration.
This mechanism is basically the same as that
found by \cite{SU06,SU08} in the case of $M_{\rm *} = 120M_\odot$ .

\begin{figure}
	\centering
	{\includegraphics[width=7cm]{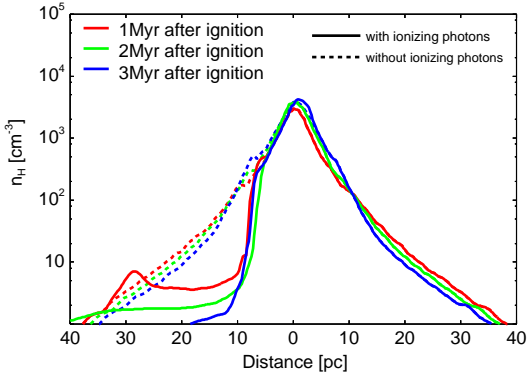}}
	\caption{Same as Fig. \ref{evo80M}, but for $M_{*} = 25M_{\odot}$.
	$D=14{\rm pc}$ is set up so that the LW band flux toward 
    the cloud core should be the same as Fig. \ref{evo80M}.}
	\label{evo25M}
\end{figure}
\begin{figure}
	\centering
	{\includegraphics[width=7cm]{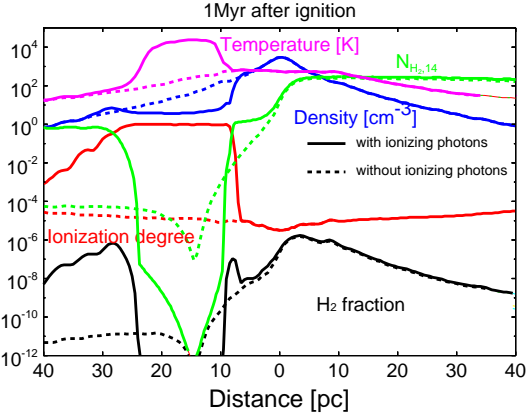}}
	\caption{Same as Fig. \ref{vpq80M}, except that the parameters are 
	$M_{*} = 25M_{\odot}$ and $D=14{\rm pc}$.}
	\label{vpq25M}
\end{figure}
In Fig. \ref{evo25M} and Fig. \ref{vpq25M}, the results
in the case of $M_{*} = 25M_{\odot}$ are shown, where
 $n_{\rm on}=10^3{\rm cm^{-3}}$.
Here, the source distance is set to be $D=14{\rm pc}$ 
so that the LW band flux toward the cloud core should 
be the same as that in the case of $M_{*} = 80M_{\odot}$,
whereas the flux of ionizing radiation is about 0.75 times weaker 
than that in the $M_{*} = 80M_{\odot}$ case.   
As shown in Fig. \ref{evo25M}, the cloud fails to collapse and 
a hydrostatic core forms, notwithstanding the presence of ionizing radiation. 
Similar to the case of $M_{*} = 80M_{\odot}$,
the H$_2$ fraction ahead of the I-front is enhanced
associated with a dense shell.
However, the H$_2$ column density of the shell is not high enough
to shield the H$_2$ dissociating photons. 
Thus, the H$_2$ fraction at the cloud core stays as low as 
$y_{\rm H_2} \approx 10^{-6}$, which is the almost same level
in the case without ionizing radiation.
Thus, in this lower stellar mass case, the ionizing radiation 
cannot suppress the negative feedback. 

%
\begin{figure}
	\centering
	{\includegraphics[width=7cm]{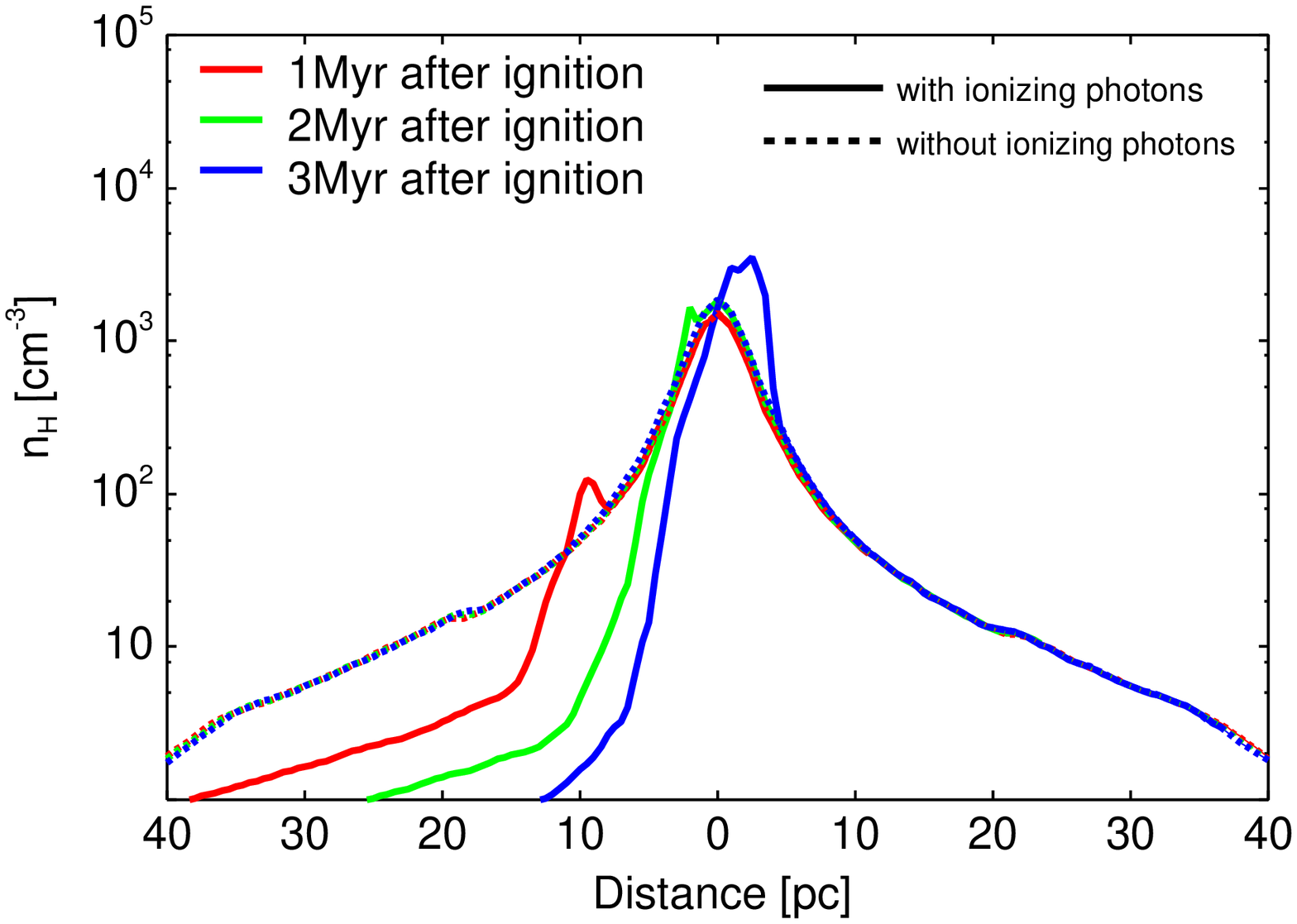}}
	\caption{Same as Fig. \ref{evo80M}, but for $T_{\rm c}\sim200$K. }
	\label{evo80Mlow}
\end{figure}
\begin{figure}
	\centering
	{\includegraphics[width=7cm]{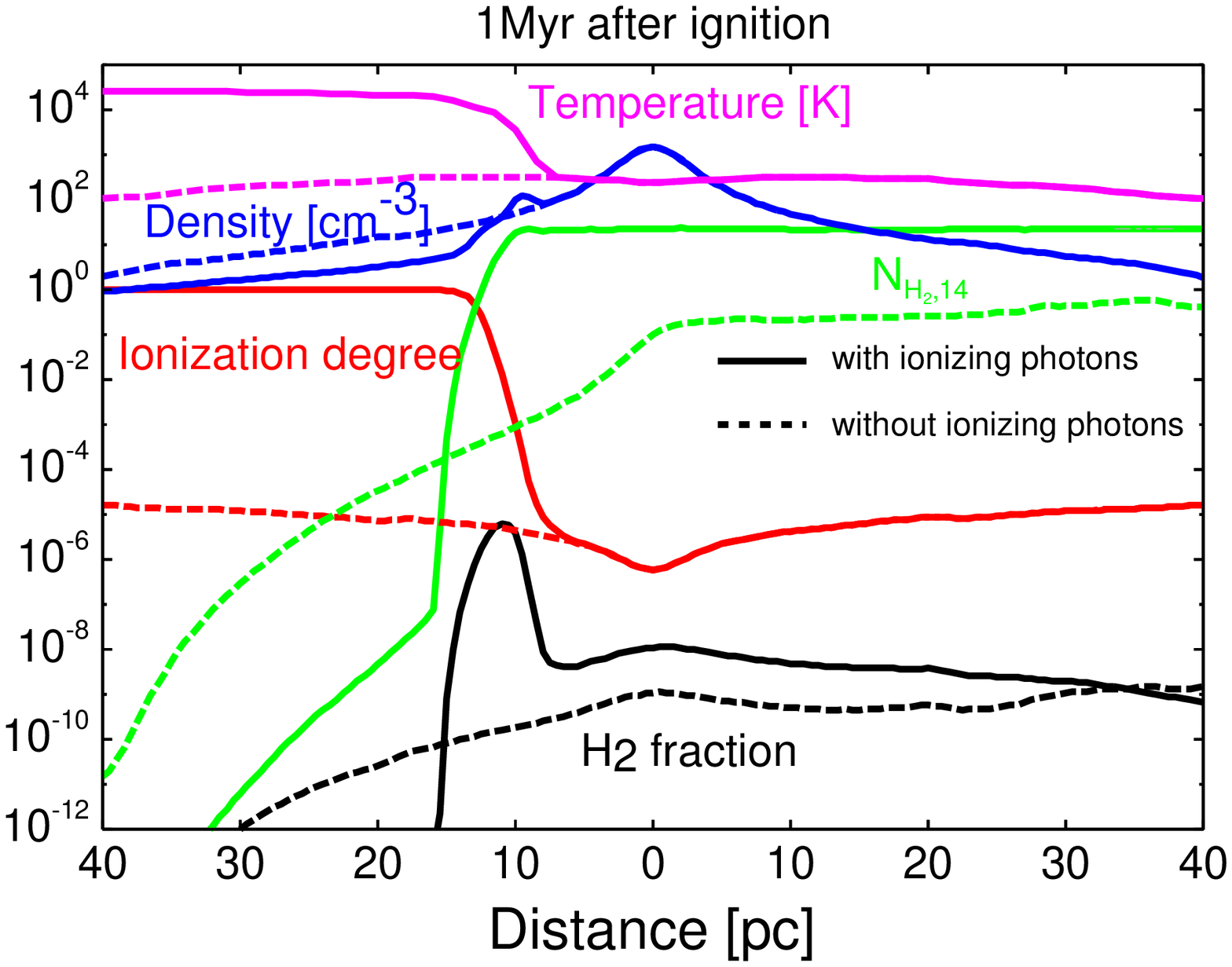}}
	\caption{Same as Fig. \ref{vpq80M}, but for $T_{\rm c}\sim200$K.}
	\label{vpq80Mlow}
\end{figure}

In Fig. \ref{evo80Mlow} and Fig. \ref{vpq80Mlow}, 
the results for the low core temperature model ($T_{\rm c}\sim 200$K) 
are shown. In these simulations, the parameters are set to be 
$M_{*}=80M_{\odot}$, $D=40$pc, and $n_{\rm on}=10^{3}{\rm cm^{-3}}$. 
As shown in Fig. \ref{evo80Mlow}, the cloud fails to collapse, 
despite the presence of ionizing radiation. 
However, the reason for the failure is different from the $M_{*}=25M_{\odot}$ case. 
It can be seen in Fig. \ref{vpq80Mlow} that the shielding effect raises 
the $\rm H_2$ column density, compared to the case with no ionizing radiation. 
In this low core temperature model, hydrogen molecules are strongly 
destroyed by the LW radiation, and the $\rm H_2$ fraction decreases to 
$y_{\rm H_2}\approx10^{-8}$. 
Since the core radius is smaller for the low core temperature,
the self-shielding for LW radiation by the core is weaker.

\section{Criteria for Radiative Feedback}
\subsection{Numerical Criteria}

\begin{figure*}
	\centering
	{\includegraphics[width=15cm]{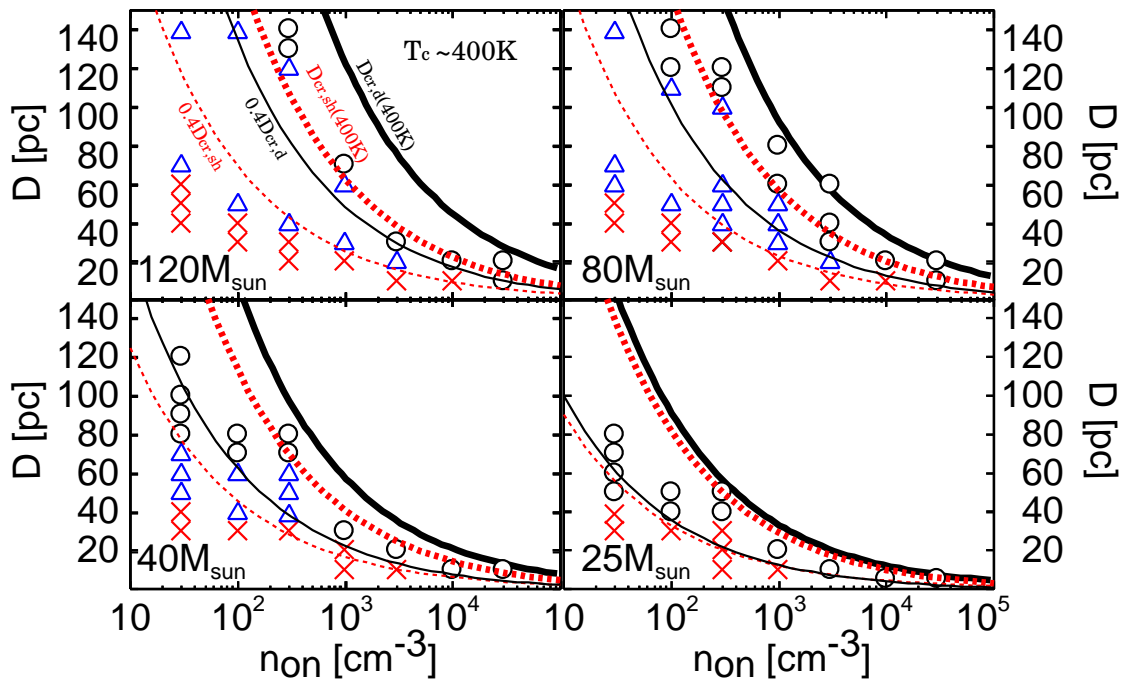}}
	\caption{Numerical results for $T_{\rm c}\sim 400$K are summarized 
    in terms of $D$ and $n_{\rm on}$. 
	From the top-left panel to the bottom-right panel, 
	each panel represents the result with $M_{*}=$ 
	$120M_{\odot}$, $80M_{\odot}$, $40M_{\odot}$, and $25M_{\odot}$. 
	Crosses denote the runs in which the clouds fail to collapse, 
	triangles represent the runs in which the clouds can collapse if the 
	ionizing radiation is included, and 
    circles represent the runs in which the clouds can collapse 
    even if the ionizing radiation is not included. 
	Thick solid lines indicate the analytic criteria $D_{\rm cr,d}$, which 
    is derived from balancing photodissociation timescale to free-fall
    timescale (see the text for the detail). 
	On the other hand, dotted lines indicate the analytic criteria $D_{\rm cr,sh}$
	including the shielding effect by the shell. 
	In each panel, $0.4D_{\rm cr,d}$ and $0.4D_{\rm cr,sh}$ are also shown by 
	a thin solid line and a thin dotted line, respectively. }
	\label{sum_t100}
\end{figure*}

In Fig. \ref{sum_t100}, the numerical results are summarized
for a high core temperature model ($T_{\rm c}\sim 300-400$K). 
In this figure, crosses denote the failed collapse, 
triangles represent the successful collapse with the aid of ionizing radiation, 
and circles represent the collapse regardless of ionizing radiation. 
As shown in this figure, in the simulation runs with $M_*\ge 40M_{\odot}$, 
the H$_2$ shell driven by ionizing radiation can allow the clouds to collapse
if the conditions for $D$ and $n_{\rm on}$ are satisfied.
However, in the case of $M_*=25M_{\odot}$, 
ionizing radiation does not help the clouds to collapse, but 
the fate of clouds is determined solely by H$_2$-dissociating
radiation. 
Hence, we conclude that the critical stellar mass below which ionizing radiation 
cannot extinguish the negative feedback by photodissociation 
is $M_* \sim 25 M_{\odot}$. 

\begin{figure*}
	\centering
	{\includegraphics[width=15cm]{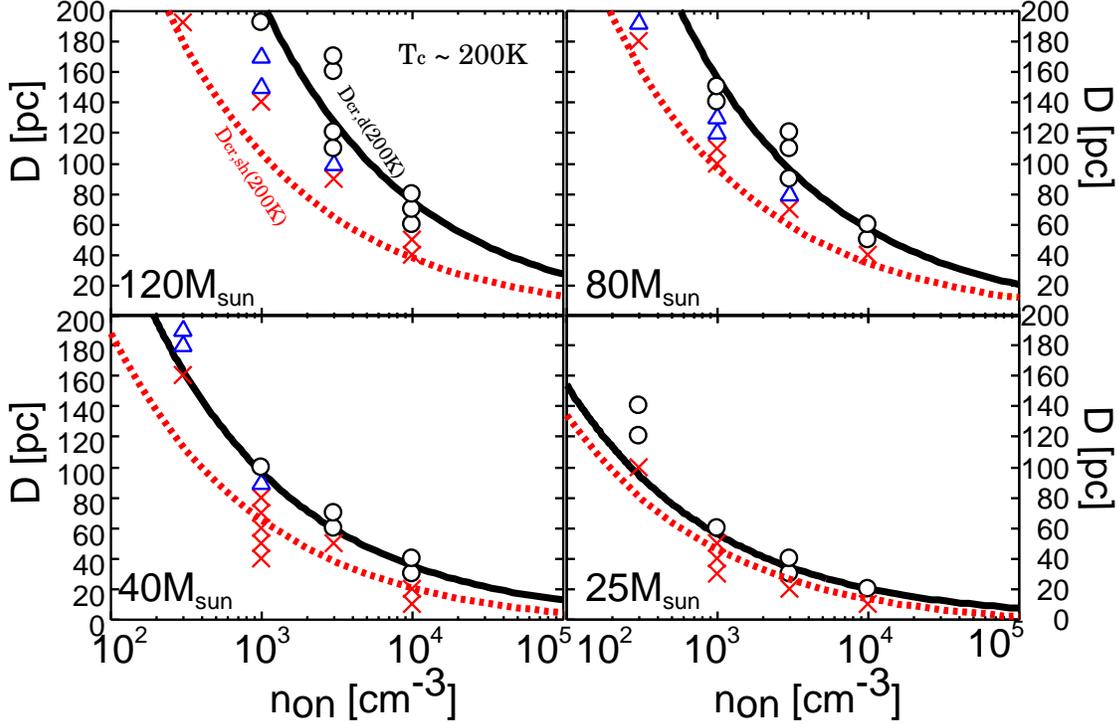}}
	\caption{Numerical results for $T_{\rm c}\sim 200$K are summarized in terms 
	of $D$ and $n_{\rm on}$. 
	From the top-left panel to the bottom-right panel, 
	each panel represents the result with $M_{*}=$ 
	$120M_{\odot}$, $80M_{\odot}$, $40M_{\odot}$, and $25M_{\odot}$. 
	Symbols have the same meanings as those in Fig. \ref{sum_t100}}
	\label{sum_t350}
\end{figure*}

In Fig. \ref{sum_t350}, the numerical results are summarized
for a low core temperature model ($T_{\rm c}\sim 200$K). 
The tendency is qualitatively the same as the results for $T_{\rm c}\sim 300-400$K. 
But, the regions of the collapse with the aid of ionizing radiation (triangles)
are obviously narrower, and in wider regions the clouds fail to collapse.
These results basically originates in the fact that the radius of cloud 
core is smaller, compared to a high core temperature model 
with $T_{\rm c}\sim 300-400$K. For the smaller core radius, H$_2$-dissociating
radiation is liable to permeate and suppress the core collapse.
However, the critical stellar mass, $M_*\sim 25 M_{\odot}$,
below which ionizing radiation 
cannot extinguish the negative feedback by photodissociation, 
is almost the same as that in the case with $T_{\rm c}\sim400$K. 
This fact means that the critical stellar mass does 
not depend sensitively on the cloud core temperature. 
As a result, we conclude that the critical stellar mass below which 
ionizing radiation is not important is $M_{*}\sim 25 M_{\odot}$.

\subsection{Analytic Criteria}
Here, we make analytic estimation of the feedback
criteria.  
\cite{Susa07} explored the photodissociation feedback of a Pop III  star 
with $120M_{\odot}$ on a neighboring prestellar core by RHD simulations 
which dose not include ionizing radiation. 
\cite{Susa07} has found  that a condition for the collapse of a neighboring core is 
approximately determined by $t_{\rm dis}=t_{\rm ff}$, where $t_{\rm dis}$ is the 
photodissociation timescale in the core and $t_{\rm ff}$ is the free-fall timescale. 
Using the condition, the critical distance $D_{\rm cr,d}$, 
below which a neighboring core fails to collapse, 
is given by 
\begin{eqnarray}
	D_{\rm cr,d}& =& 147 {\rm pc}
	\left(\frac{L_{\rm LW}}{5\times 10^{23}{\rm erg\;s^{-1}}}\right)^{\frac{1}{2}}
	\left(\frac{n_{\rm c}}{10^{3}{\rm cm^{-3}}}\right)^{-\frac{7}{16}} \nonumber \\
	&&\times \left(\frac{T_{\rm c}}{300{\rm K}}\right)^{-\frac{3}{4}},  \label{Dcrd}
\end{eqnarray}
where $L_{\rm LW}$, $n_{\rm c}$, and $T_{\rm c}$ are the 
LW luminosity of source star, the number density of core, and 
the temperature of core, respectively. 
This equation involves the self-shielding effect by the core. 
The dependence on the core temperature basically 
originates in the core radius ($\propto T_{\rm c}^{1/2}$) 
and the $\rm H_2$ formation rate in the core ($\propto T_{\rm c}$). 
Hence, the self-shielding effect is weaker for the lower core temperature. 
As argued in \cite{Susa07}, 
the boundary between the collapses regardless of ionizing radiation (circles) 
and with the aid of ionizing radiation (triangles) in Fig. \ref{sum_t100} 
can be roughly explained by 
$D_{\rm cr,d}$ in the case with $M_*=120M_{\odot}$. 
In addition, as shown in Figs. \ref{sum_t100} and \ref{sum_t350},
$D_{\rm cr,d}$ gives a good estimate 
for less massive source star cases. 
However, as shown in Figs. \ref{sum_t100}, the boundary for the 
high core temperature model is slightly lower than this analytic estimate. 
This disagreement can be understand by the dynamical effect of the
collapsing clouds \citep{Susa07}.
The actual dynamical contraction is faster for the high core temperature 
model, since the ratio of gravitational energy to internal energy is higher
($|W|/U \approx 4$) as described in \S\ref{code}.
Then, the $\rm H_2$ fraction in the core recovers rapidly,
during the adiabatic compression phase. Hence,
the core can keep collapsing, even if the photodissociation timescale
is shorter than the free-fall timescale
($t_{\rm dis}< t_{\rm ff}$) when the cloud irradiated by UV.
As a result, the criterion is reduced to $f_{\rm dyn} D_{\rm cr,d}$ 
by a dynamical factor $f_{\rm dyn}$.
Compared to the numerical results, 
we find $f_{\rm dyn} \approx 0.4$ for the high core temperature model.
On the other hand, for the low core temperature model, 
the dynamical effect is not so strong because of $|W|/U \approx 2$,
and therefore $f_{\rm dyn} \approx 1$. 

Furthermore, if the ionizing radiation is included,
we should incorporate the shielding effect by an $\rm H_2$ shell. 
Here, we derive a new criterion including the this effect. 
Since a cloud collapses in a self-similar fashion before UV irradiation, 
the density profile of outer envelope in the cloud is expressed as 
\begin{equation}
	n(r) = n_{\rm c}\left(\frac{r_{\rm c}}{r}\right)^2, 
\end {equation}
where $r_{\rm c}$ is the core radius which roughly corresponds to 
the Jeans scale; 
\begin{equation}
	r_{\rm c} = \frac{1}{2}\sqrt{\frac{\pi k_B T_{\rm c}}{Gm_p^2n_{\rm c}}}, 
\end{equation} 
where $k_{\rm B}$ denotes the Boltzmann constant 
and $m_{\rm p}$ denotes the proton mass. 
Assuming that the thickness of the $\rm H_2$ shell is determined by 
the amount of ionized gas in the envelope 
and the $\rm H_{2}$ fraction in the shell is constant,   
the $\rm H_2$ column density of the shell $N_{\rm H_2,sh}$ is given by 
\begin{equation}
	N_{\rm H_2,sh}=\int_D^{D_{\rm sh}}y_{\rm H_2, sh} n(r) dr
	=y_{\rm H_2,sh}n_{\rm c}r_{\rm c} ^2 \frac{D-D_{\rm sh}}{DD_{\rm sh}}, \label{NH2sh1}
\end{equation}
where $D_{\rm sh}$, and $y_{\rm H_2,sh}$ 
are the distance between the cloud core and the ${\rm H_2}$ shell, 
and the $\rm H_2$ fraction in the shell, respectively.   
Here, $D_{\rm sh}$ is set to be the position where the number of recombination 
per unit time in the ionized region around a source star balances with the number 
rate of incident ionizing photons, since the $\rm H_2$ shell appears ahead 
of ionization front. 
Hence, $D_{\rm sh}$ satisfies  
\begin{eqnarray}
	\frac{N_{\rm ion}\pi D_{\rm sh}^2}{4\pi (D-D_{\rm sh})^2 }
	&=&2\pi \alpha_{\rm B}\int^{D_{\rm sh}}_{D}n(r)^2r^2dr \nonumber \\
	&=&2\pi \alpha_{\rm B }n_c^2 r_c^4\frac{D-D_{\rm sh}}{DD_{\rm sh}}, \label{Nion}
\end{eqnarray}
where $\alpha_{\rm B}$ is the recombination coefficient 
to all excited levels of hydrogen.  
Using equation (\ref{NH2sh1}) and (\ref{Nion}), we obtain 
\begin{equation}
	N_{\rm H_2,sh}=y_{\rm H_2,sh} n_{\rm c}^{\frac{1}{3}}r_{\rm c}^{\frac{2}{3}}
	D^{-\frac{2}{3}}\left(\frac{N_{\rm ion}}{8\pi \alpha_{\rm B}}\right)^{\frac{1}{3}}
	 \label{NH2sh2}
\end{equation}
Because of the intense LW radiation, the $\rm H_2$ at the shell is in chemical 
equilibrium. Therefore, $y_{\rm H_2,sh}$ is given by
\begin{equation}
	y_{\rm H_2,sh}=\frac{n(D_{\rm sh})y_{\rm e,sh} k_{\rm H^-}}{k_{\rm dis}}, 
\end{equation}
where $y_{\rm e,sh}$ is the electron fraction at the $\rm H_2$ shell 
and $k_{\rm H^-}$  is the reaction rate of $\rm H^-$ process. 
In this case, we should consider the self-shielding effect by the shell itself. 
As a result, these rates are 
\begin{equation}
	k_{\rm H^-} = 1.0\times 10^{-18} T_{\rm sh} {\rm cm^{-3}s^{-1}}, 
\end{equation}
\begin{equation}
	k_{\rm dis} = 1.13 \times 10^8 F_{\rm LW,sh} 
	f_{\rm s}\left(\frac{N_{\rm H_2,sh}/2}{10^{14}\rm cm^{-2}}\right) \rm s^{-1}, \label{kdis}
\end{equation}
where $T_{\rm sh}$ and $F_{\rm LW,sh}$ are the temperature at the shell, and 
the LW flux from the star in the absence of shielding effect,  
$F_{\rm LW,sh}=L_{\rm LW}/4\pi(D-D_{\rm sh})^2$. 
In addition, $f_{\rm s}$ is the self-shielding function given by
(\ref{shield-function}).  
Combining equations (\ref{NH2sh2})-(\ref{kdis}) 
with assumption of $y_{\rm e,sh}=0.05$ and $T_{\rm sh}=2000$K 
as shown in the present numerical results, we have 
 \begin{eqnarray}
	y_{\rm H_2,sh}
	&=&1.0\times 10^{-6}\left(\frac{N_{\rm ion}}{10^{50}{\rm s^{-1}}}\right)^{\frac{11}{3}}
	\left(\frac{L_{\rm LW}}{5\times10^{23}{\rm erg\;s^{-1}}}\right)^{-4}\nonumber \\
	&&\times \left(\frac{T_{\rm c}}{300{\rm K}}\right)^{-\frac{1}{3}}
	\left(\frac{D}{40{\rm pc}}\right)^{\frac{2}{3}},
\end{eqnarray}
\begin{eqnarray}
	N_{\rm H_2,sh}&=&5.8\times10^{14}\left(\frac{N_{\rm ion}}{10^{50}{\rm s^{-1}}}\right)^4
	\nonumber \\
	&& \times \left(\frac{L_{\rm LW}}{5\times 10^{23}{\rm erg\;s^{-1}}}\right)^{-4}{\rm cm^{-2}}. 
	\label{NH2sh3}
\end{eqnarray}
Notice that $N_{\rm H_2,sh}$ 
is determined solely by the ratio of $N_{\rm ion}$ to 
$L_{\rm LW}$, and strongly depends on the ratio. 
In the above numerical results, it is shown that the critical stellar mass does 
not depend sensitively on the cloud core temperature. 
This fact is consistent with equation (\ref{NH2sh3}), in which 
the $\rm H_2$ column density of shell is independent of 
the core temperature $T_{\rm c}$. 

Multiplying $L_{\rm LW}$ in equation (\ref{Dcrd}) by 
$f_{\rm s,sh}\equiv f_{\rm s}\left(\frac{N_{\rm H_2,sh}}{10^{14}{\rm cm^{-2}}}\right)$, 
we obtain the critical distance as  
\begin{eqnarray}
	D_{\rm cr,sh}& =& 147 {\rm pc}
	\left(\frac{L_{\rm LW}f_{\rm s, sh}}
	{5\times 10^{23}{\rm erg\;s^{-1}}}\right)^{\frac{1}{2}}
	\left(\frac{n_{\rm c}}{10^{3}{\rm cm^{-3}}}\right)^{-\frac{7}{16}} \nonumber \\
	&&\times \left(\frac{T_{\rm c}}{300{\rm K}}\right)^{-\frac{3}{4}},  \label{Dcrsh}
\end{eqnarray}
in which both shielding effects by the core and the $\rm H_2$ shell are taken into account. 
In particular, when $N_{\rm H_2,sh}>10^{14} \rm cm^{-2}$, 
the critical distance can be expressed as 
\begin{eqnarray}
	D_{\rm cr,sh} &=&78.8 {\rm pc}
	\left(\frac{L_{\rm LW}}
	{5\times 10^{23}{\rm erg\;s^{-1}}}\right)^{2}\left(\frac{N_{\rm ion}}{10^{50}{\rm s^{-1}}}
	\right)^{-\frac{3}{2}}\nonumber \\
	&&\times \left(\frac{n_{\rm c}}{10^{3}{\rm cm^{-3}}}\right)^{-\frac{7}{16}}
	\left(\frac{T_{\rm c}}{300{\rm K}}\right)^{-\frac{3}{4}}. 
\end{eqnarray}
In Fig. \ref{sum_t100} and Fig. \ref{sum_t350},
triangles are the collapse with the aid of an $\rm H_2$ shell.
Therefore, the boundary between the triangles and the crosses 
should be compared with $D_{\rm cr, sh}$. 
According to equation (\ref{NH2sh3}), the shielding effect by the shell becomes 
weaker according as $N_{\rm ion}/L_{\rm LW}$ decreases. 
This indicates that $D_{\rm cr,sh}$ approaches $D_{\rm cr,d}$ 
as the mass of source star becomes lower. 
As shown in Fig. \ref{sum_t100}, $D_{\rm cr,sh}$ is the almost same 
as $D_{\rm cr,d}$ in the case with $M_*=25M_{\odot}$. 
This result originate in the 
strong dependence of the shell $\rm H_2$ column density on 
$N_{\rm ion}/L_{\rm LW}$ (see equation \ref{NH2sh3}). 
As shown in Figs. \ref{sum_t100} and \ref{sum_t350},
$D_{\rm cr,sh}$ gives a qualitatively good estimate for
the collapse with the aid of an $\rm H_2$ shell.
However, the boundary in Figs. \ref{sum_t100} is 
slightly lower than this analytic estimate.
For the same reason as in the case of $D_{\rm cr,d}$, 
the dynamical effect is more
prominent for the high core temperature model. 
In this case, $f_{\rm dyn} D_{\rm cr,sh}$ with $f_{\rm dyn} \approx 0.4$ 
provides a more appropriate criterion. 
On the other hand, for the low core temperature model, 
$f_{\rm dyn} \approx 1$ gives a plausible criterion.

\section{Conclusions and Discussion}

We have carried out RHD simulations to investigate the impact of UV radiation 
from a Pop III star on nearby collapsing cores. 
In particular, our attention has been paid to the dependence of UV feedback 
on the mass of Pop III star. 
The radiation hydrodynamic evolution of cloud core is determined
by not only $\rm H_2$-dissociating radiation but also
ionizing radiation. 
As a result, we have found the critical stellar mass $M_{*}\approx 25 M_{\odot}$,
above which an $\rm H_2$ shell ahead of ionizing front can 
help clouds to collapse. 
Owing to the fact that $\rm H_2$-dissociating radiation becomes 
predominant for less massive source stars, the critical distance 
for the collapse of a neighboring core 
does not so strongly depend on the mass of source star. 
Also, we have derived analytically the feedback criterion,
$f_{\rm dyn} D_{\rm cr,sh}$, where $D_{\rm cr,sh}$ is given by (\ref{Dcrsh})
and $f_{\rm dyn}$ is a dynamical factor dependent on the 
the ratio of gravitational energy $W$ to internal energy $U$ of collapsing
cloud. 
We have found $f_{\rm dyn} \approx 0.4$ for $|W|/U \approx 4$,
and $f_{\rm dyn} \approx 1$ for $|W|/U \approx 2$. 
Since $f_{\rm dyn}$ is dependent on $|W|/U$,  
a dark matter (DM) halo can influence the feedback criterion to a certain degree.
In order to assess the effects of DM, we have calculated several models
with a static NFW-type dark matter halo potential 
\citep{NFW97} with $M_{\rm vir}=4.15\times 10^5 M_{\odot}$ 
and $r_{\rm vir}=160{\rm pc}$.
In these runs, the ratios of DM mass ($M_{\rm DM}$) to baryonic mass 
($M_{\rm b}$) at the central regions of $r<10$pc are  
$M_{\rm DM}/M_{\rm b} \simeq 0.3$ for $n_{\rm on}=10^3{\rm cm^{-3}}$, 
and $M_{\rm DM}/M_{\rm b} \simeq 1$ for $n_{\rm on}=10^2{\rm cm^{-3}}$. 
As a result, we have found 
that the feedback criterion in the form of $f_{\rm dyn} D_{\rm cr,sh}$ turns out to be 
still valid, and $f_{\rm dyn}$ becomes smaller by a factor of 1.2 
for $n_{\rm on}=10^3{\rm cm^{-3}}$ and 
by a factor of 2 for $n_{\rm on}=10^2{\rm cm^{-3}}$. 
Therefore, our main results are not changed so much by including DM. 
Note that the DM density evolution is not treated consistently with the gas dynamics
in these simulations. 
If the DM dynamics is solved with the evolution of gas clouds, 
the evolutionary path of core temperature might be changed. 
Hence, for a more quantitative argument, the self-consistent treatment
of dark matter would be requisite. 

In this paper, we have not considered the lifetime of source stars. 
The lifetime of Pop III star is 
$2.5 \times 10^6$yr for 120$M_\odot$,
$3.0 \times 10^6$yr for 80$M_\odot$, 
$3.9 \times 10^6$yr for 40$M_\odot$,
and 
$6.5 \times 10^6$yr for 25$M_\odot$ \citep{Schaerer02}.
If the lifetime of source star is shorter than the free-fall time
determined by $n_{\rm on}$,
the feedback may be significantly changed before the cloud
collapse. The density in which the free-fall time equals 
the stellar lifetime is  
$n_{\rm on} =419 {\rm cm}^{-3}$ for 120$M_\odot$,
$n_{\rm on} =293 {\rm cm}^{-3}$ for 80$M_\odot$, 
$n_{\rm on} =178 {\rm cm}^{-3}$ for 40$M_\odot$,
and 
$n_{\rm on} =64 {\rm cm}^{-3}$ for 25$M_\odot$.
Below these densities, arguments including the effects from
the stellar lifetime are requisite. 

The fate of Pop III stars depends on the mass \citep{Heger02,Heger03}.
Pop III stars with 120$M_\odot$ or 80$M_\odot$ may result in
direct collapse to black holes (BHs), while 
those with 40$M_\odot$ or 25$M_\odot$ may undergo Type II supernova
explosions.
In the case of direct BH formation, UV source disappears
abruptly, and then already-formed $\rm H_2$ molecules can promote
the collapse of cloud cores \citep[e.g.,][]{Nagakura05,JB06,GB06,Yoshida07a}. 
In the case of Type II SN explosions, shock-driven hydrodynamic
feedbacks could be significant \citep{Mori02,Bromm03,KY05,Greif07}.

\section*{Acknowledgements}
Numerical simulations have been performed with computational facilities 
at Center for Computational Sciences in University of Tsukuba. 
This work was supported in part by the {\it FIRST} project based on
Grants-in-Aid for Specially Promoted Research by 
MEXT (16002003) and Grant-in-Aid for Scientific 
Research (S) by JSPS  (20224002), and also 
in part by Inamori Research Foundation.

\label{lastpage}


\begin{thebibliography}{99}
\bibitem[\protect\citeauthoryear{Abel, Bryan \& Norman}{2000}]{Abel00} 
Abel T., Bryan G. L., Norman M. L., 2000, ApJ, 540, 39
\bibitem[\protect\citeauthoryear{Ahn \& Shapiro}{2007}]{AS07} 
Ahn K., Shapiro P.~R., 2007, MNRAS, 375, 881 
\bibitem[\protect\citeauthoryear{Bromm, Coppi \& Larson}{2002}]{Bromm02}
Bromm V., Coppi P. S., Larson R. B., 2002, ApJ, 564, 23
\bibitem[Bromm et al.(2003)]{Bromm03} 
Bromm V., Yoshida N., Hernquist L., 2003, ApJ, 596, L135 
\bibitem[\protect\citeauthoryear{Cen}{2003}]{Cen03} 
Cen R., 2003, ApJ, 591, L5 
\bibitem[\protect\citeauthoryear{Ciardi et al.}{2001}]{Ciardi01} 
Ciardi B., Ferrara A., Marri S., Raimondo G., 2001, MNRAS, 324, 381
\bibitem[\protect\citeauthoryear{Draine \& Bertoldi}{1996}]{DB96}
Draine B. T., Bertoldi F., 1996, ApJ, 468, 269
\bibitem[Fuller \& Couchman(2000)]{Fuller00} 
Fuller T.~M., Couchman H.~M.~P., 2000, ApJ 544, 6 
\bibitem[\protect\citeauthoryear{Galli \& Palla}{1998}]{GP98} 
Galli D., Palla F., 1998, A \& A, 335, 403
\bibitem[\protect\citeauthoryear{Glover \& Brand}{2001}]{GB01} 
Glover S. C. O., Brand P. W. J. L., 2001, MNRAS, 321, 385 
\bibitem[\protect\citeauthoryear{Gnedin}{2000}]{Gnedin00} 
Gnedin N.~Y., 2000, ApJ, 535, 530 
\bibitem[\protect\citeauthoryear{Greif \& Bromm}{2006}]{GB06}
Greif T.~H., Bromm V., 2006, MNRAS,373, 128 
\bibitem[Greif et al.(2007)]{Greif07} 
Greif T.~H., Johnson J.~L., Bromm V., Klessen R.~S.,\ 2007, ApJ, 670, 1 
\bibitem[\protect\citeauthoryear{Haiman, Rees \& Loeb}{1997}]{Haiman97} 
Haiman Rees, M. J., Loeb A., 1997, ApJ, 476, 458
\bibitem[\protect\citeauthoryear{Heger et al.}{2003}]{Heger03} 
Heger A., Fryer C.~L., Woosley S.~E., Langer N., Hartmann D.~H., 2003, ApJ, 
591, 288 
\bibitem[\protect\citeauthoryear{Heger \& Woosley}{2002}]{Heger02} 
Heger A., Woosley S.~E., 2002, ApJ, 567, 532 
\bibitem[\protect\citeauthoryear{Iwamoto et al.}{2005}]{Iwamoto05} 
Iwamoto N., Umeda H., Tominaga N., Nomoto K., Maeda K. 2005, 
Science, 309, 451
\bibitem[\protect\citeauthoryear{Johnson \& Bromm}{2006}]{JB06} 
Johnson J.~L., Bromm V., 2006, MNRAS, 366, 247 
\bibitem[Johnson et al.(2008)]{Johnson08} 
Johnson J.~L., Greif T.~H., Bromm, V., 2008, MNRAS, 694 
\bibitem[Kang \& Shapiro (1992)]{KS92}
Kang H., Shapiro P., ApJ, 386, 432
\bibitem[\protect\citeauthoryear{Kitayama et al.}{2001}]{Kitayama01}
Kitayama T., Susa H.,Umemura M., Ikeuchi S., 2001, MNRAS, 326, 1353
\bibitem[Kitayama \& Yoshida(2005)]{KY05} 
Kitayama T., Yoshida N.,\ 2005, ApJ, 630, 675 
\bibitem[\protect\citeauthoryear{Machacek, Bryan \& Abel}{2001}]{Macha01}
Machacek M.E., Bryan G. L., Abel T., \ 2001, ApJ, 548, 509
\bibitem[Mori et al.(2002)]{Mori02} 
Mori M., Ferrara A., Madau P.,\ 2002, ApJ, 571, 40 
\bibitem[\protect\citeauthoryear{Nagakura \& Omukai}{2005}]{Nagakura05} 
Nagakura T., Omukai K., 2005, MNRAS, 364, 1378 
\bibitem[\protect\citeauthoryear{Nakamoto, Umemura \& Susa}{2001}]{b3}
Nakamoto T., Umemura M., Susa H., 2001, MNRAS, 321, 593
\bibitem[\protect\citeauthoryear{Nakamura \& Umemura}{2001}]{NU01} 
Nakamura F., Umemura M., 2001, ApJ, 548, 19 
\bibitem[\protect\citeauthoryear{Nakamura \& Umemura}{2002}]{NU02} 
Nakamura F., Umemura M., 2002, ApJ, 569, 549 
\bibitem[\protect\citeauthoryear{Navarro, Frenk \& White}{1997}]{NFW97} 
Navarro J. F., Frenk C. S., \& White S. D. M., 1997, ApJ, 490, 493
\bibitem[\protect\citeauthoryear{Oh \& Haiman}{2002}]{OH02}
Oh S. P., Haiman Z., 2002, ApJ, 569, 558
\bibitem[\protect\citeauthoryear{Omukai \& Nishi}{1999}]{ON99}
Omukai K., Nishi R., 1999, ApJ, 518, 64
\bibitem[\protect\citeauthoryear{O'Shea \& Norman}{2007}]{Oshea07}
O'Shea B.~W., Norman M.~L., 2007, ApJ, 654, 66 
\bibitem[\protect\citeauthoryear{O'Shea \& Norman}{2008}]{Oshea08} 
O'Shea B.~W., Norman M.~L., 2008, ApJ, 673, 14 
\bibitem[Ricotti, Gnedin, \& Shull(2001)]{Ricotti01}
Ricotti M., Gnedin N.~Y., Shull, M., 2001, ApJ, 560, 580
\bibitem[\protect\citeauthoryear{Schaerer}{2002}]{Schaerer02}
Schaerer D., 2002, A\&A, 382, 28
\bibitem[\protect\citeauthoryear{Shapiro \& Kang}{1987}]{SK87}
Shapiro P. R., Kang H., 1987, ApJ, 318, 32
\bibitem[\protect\citeauthoryear{Sokasian et al.}{2004}]{Sokasian04} 
Sokasian A., Yoshida N., Abel T., Hernquist L., Springel V., 2004, MNRAS, 350, 47 
\bibitem[\protect\citeauthoryear{Susa}{2006}]{Susa06} 
Susa H., 2006, PASJ, 58, 455
\bibitem[\protect\citeauthoryear{Susa}{2007}]{Susa07} 
Susa H., 2007, ApJ, 659, 908
\bibitem[\protect\citeauthoryear{Susa et al.}{1998}]{Susa98} 
Susa H., Uehara H., Nishi R., Yamada M., 1998, PThPh, 100, 63 
\bibitem[\protect\citeauthoryear{Susa \& Umemura}{2004a}]{SU04a} 
Susa H., Umemura M., 2004, ApJ, 600, 1 
\bibitem[\protect\citeauthoryear{Susa \& Umemura}{2004b}]{SU04b} 
Susa H., Umemura M., 2004, ApJ, 610, L5 
\bibitem[\protect\citeauthoryear{Susa \& Umemura}{2006}]{SU06} 
Susa H., Umemura M., 2006, ApJ, 645, L93
\bibitem[\protect\citeauthoryear{Susa, Umemura \& Hasegawa}{2008}]{SU08} 
Susa H., Umemura M., Hasegawa K., 2008, in preparation 
\bibitem[\protect\citeauthoryear{Tajiri \& Umemura}{1998}]{TU98}
Tajiri Y., Umemura M., 1998, ApJ, 502, 59
\bibitem[Tegmark et al.(1997)]{Tegmark97} 
Tegmark M., Silk J., Rees M.~J., Blanchard A., Abel T., Palla F.,\ 1997, ApJ, 474, 1
\bibitem[Tornatore et al.(2007)]{Tornatore07} 
Tornatore L., Ferrara A., Schneider R., 2007, MNRAS, 382, 945 
\bibitem[\protect\citeauthoryear{Thoul \& Weinberg}{1996}]{TW96}
Thoul A. A., Weinberg D. H., 1996, ApJ, 465, 608
\bibitem[\protect\citeauthoryear{Uehara \& Inutsuka}{2000}]{UI00}
Ueharam H., Inutsuka S., 2000, ApJ, 531, L91
\bibitem[\protect\citeauthoryear{Umemura \& Ikeuchi}{1984}]{UI84} 
Umemura M., Ikeuchi S., 1984, PThPh, 72, 47 
\bibitem[Umemura et al.(2007)]{Umemura07} 
Umemura M., Susa H., Suwa T., Sato D., FIRST Project Team, 2007,
{\it First Stars III} (Eds. O'Shea, B.W., Heger, A., \& Abel, T.), 386
\bibitem[\protect\citeauthoryear{Umeda \& Nomoto}{2003}]{Umeda03} 
Umeda H., Nomoto K., 2003, Nature, 422, 871
\bibitem[Whalen et al.(2008)]{Whalen08} 
Whalen D., O'Shea B.~W., Smidt J., Norman M.~L.,\ 2008, ApJ, 679, 925 
\bibitem[Wise \& Abel (2008)]{WA08} 
Wise J. H., Ablel T., 2008, ApJ, 685, 40 
\bibitem[\protect\citeauthoryear{Yoshida et al.}{2003}]{Yoshida03}
Yoshida N., Abel T., Hernquist L., Sugiyama N., 2003, ApJ, 592, 645
\bibitem[\protect\citeauthoryear{Yoshida et al.}{2006}]{Yoshida06}
Yoshida N., Omukai K., Hernquist L., Abel, T., 2006, ApJ, 652, 6
\bibitem[\protect\citeauthoryear{Yoshida, Omukai \& Hernquist}{2007a}]{Yoshida07a}
Yoshida N., Omukai K., Hernquist L., 2007, ApJ, 667, L117
\bibitem[\protect\citeauthoryear{Yoshida et al.}{2007b}]{Yoshida07b} 
Yoshida N., Oh S.~P., Kitayama T., Hernquist L., 2007, ApJ, 663, 687 
\end{thebibliography}
\end{document}